\documentclass[a4paper,12pt]{article}

\usepackage[T1]{fontenc}
\usepackage[ansinew]{inputenc}
\usepackage{fancyhdr}
\usepackage{amstext}
\usepackage{epsfig}
\usepackage{graphicx}
\usepackage{latexsym}
\usepackage{dcolumn}
\usepackage{citesort}
\usepackage{bm}
\usepackage{rotating}

\newcommand{\Hbar}{$\overline{\text H}$}
\newcommand{\pbar}{$\overline{\text p}$}

\begin{document}

\title{FLAIR:\\
A Facility for Low-energy Antiproton\\ and Ion Research}
\author{E. Widmann\thanks{on behalf of the FLAIR community
    http://www-linux.gsi.de/$\sim$flair/ }\\
\\ {\em Department of Physics, University of Tokyo}\\
 {\em 7-3-1 Hongo, Bunkyo-ku, Tokyo 113-0033, Japan}}

\date{ }
\maketitle

\begin{abstract}
The future accelerator facility for beams of ions and antiprotons
at Darmstadt will provide antiproton beams of intensities that are
two orders of magnitude higher than currently available. Within
the foreseen scheme, antiprotons can be decelerated to 30 MeV. The
low-energy antiproton community has recently formed a users group
to make use of this opportunity to create a next-generation
low-energy antiproton facility called FLAIR, which will be able to
provide cooled antiproton beams well below 100 keV kinetic energy.
This talk gives an overview of the layout and physics program of
the proposed facility.
\end{abstract}

\section{Introduction}

Low-energy antiproton physics is currently being done at the
Antiproton Decelerator (AD) of CERN, Geneva. Due to the low
intensity ($\sim 10^5$ \pbar /s) and the availability of only
pulsed extraction, the physics program is limited to the
spectroscopy of antiprotonic atoms and antihydrogen formed in
charged particle traps or by stopping antiprotons in low-density
gas targets. Furthermore, the output energy of the AD (5 MeV
kinetic energy) is still significantly higher than the $< 100$ keV
energy best suited for these experiments.

At the future accelerator facility for beams of ions and
antiprotons at Darmstadt it will become possible to create a
next-generation low-energy antiproton facility to overcome these
limitations by providing cooled beams at higher intensities and a
factor 100 lower energy. In addition the new facility should have
the possibility of slow ({\em i.~e.} continuous) extraction, which
will allow nuclear/particle physics type experiments requiring
coincidence measurements to be performed.

Recently, a letter of intent has been submitted to GSI, Darmstadt,
for a facility called  FLAIR (Facility for Low-energy Antiproton
and Ion Research) \cite{FLAIR-LOI} that is described in the
following. It consists of two storage rings, a magnetic (LSR) and
an electrostatic (USR) one, and a universal trap facility
(HITRAP), cf. Fig.~\ref{fig:facility}. These components of the
facility can provide stored as well as fast and slow extracted
cooled beams at energies between 30 MeV and 300 keV (LSR), between
300 keV and 20 keV (USR), and cooled particles at rest or at
ultra-low (eV--keV) energies (HITRAP). This will allow a large
variety of new experiments to be performed, as described in
Sec.~\ref{sec:physics}. Among the unique experiments only possible
at such a facility are nuclear physics studies using antiprotons
as a hadronic probe to investigate the structure of nuclei,
including radioactive isotopes produced at the future facility,
and many atomic-collision type experiments with internal targets
in both storage rings with effective intensities as large as
10$^{10}$ \pbar /s. An important synergetic aspect is that the
whole structure will also be used to study highly charged ions,
including storing, cooling (LSR, USR) and trapping them in Penning
traps like HITRAP and investigating them in a dedicated area for
heavy ions.

\section{Layout and performance of the facility}

\begin{figure}[t]
\begin{center}
\epsfig{file=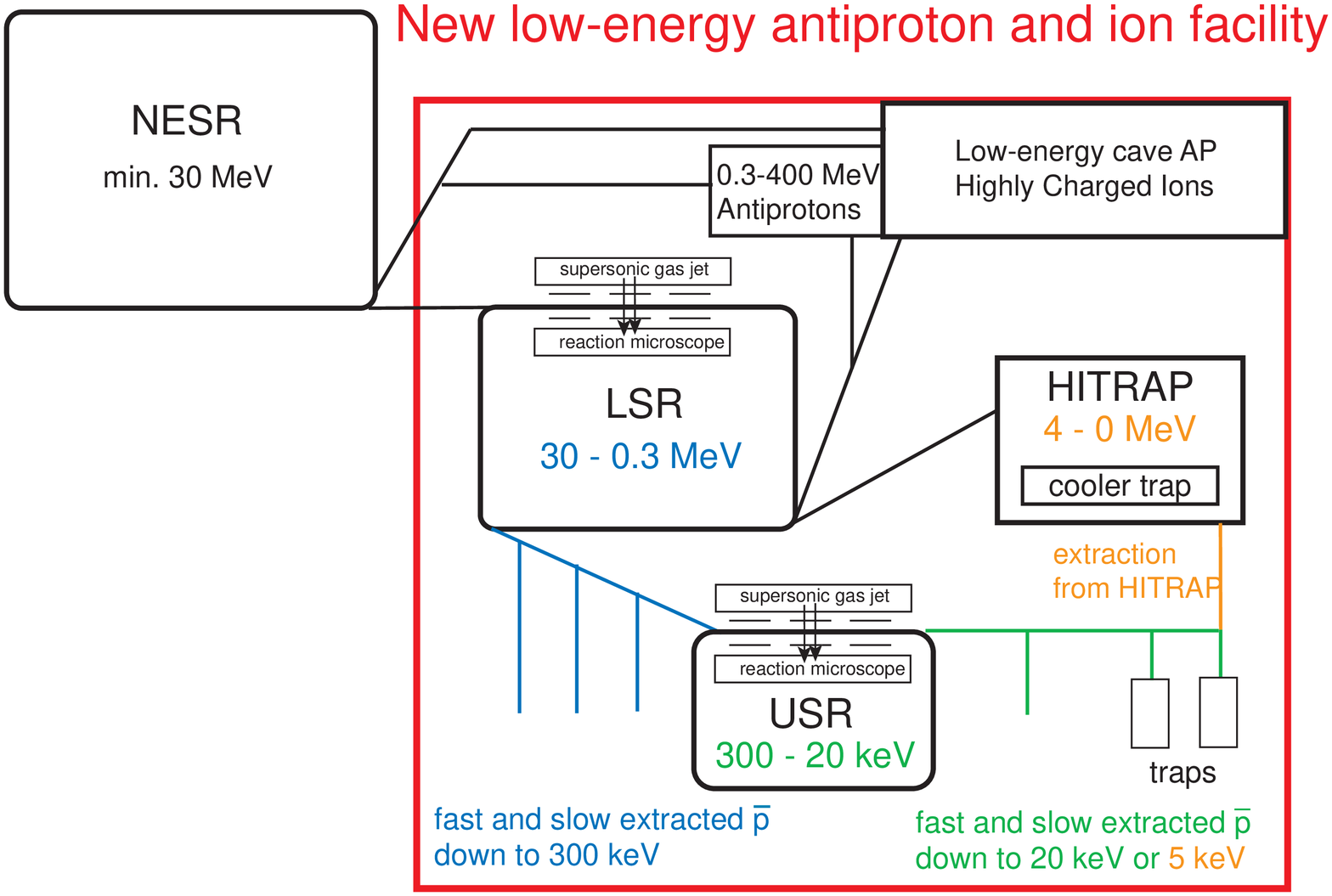,width=0.9\textwidth}
\caption{Layout of the low-energy antiproton and ion facility.
NESR: new experimental storage ring (part of the approved
project). LSR: intermediate storage ring which decelerates \pbar\
to 300 keV. USR: electrostatic ultra-low energy storage ring. Both
rings use electron cooling to provide low-emittance beams and
include internal gas jet targets for atomic collision studies.
HITRAP: trap facility for efficient deceleration and cooling of
\pbar s from 4 MeV to rest.} \label{fig:facility}
\end{center}
\end{figure}

The key features of the proposed facility at Darmstadt (cf.
Fig.~\ref{fig:facility}) are:

\begin{itemize}
    \item {\bf High-brightness, high-intensity, low-energy antiproton
          beams.}
 \begin{itemize}
    \item High antiproton intensity due to {\bf accumulation}.
    \item Cooled \pbar\ beams down to {\bf 300 keV} using LSR.
    \item Electrostatic storage ring (USR) for atomic collision
          experiments and  deceleration and cooling to {\bf
          20 keV}.
    \item HITRAP for efficient deceleration of \pbar s from 4 MeV
          to {\bf rest} and extraction of \pbar s from a cooler trap at
          {\bf keV} energies.
  \end{itemize}
    \item Both {\bf slow and fast} extraction from LSR and USR at
    energies between 30 MeV and 20 keV.
\end{itemize}

\begin{figure}[h]
\begin{center}
\includegraphics[height=0.95\textwidth,angle=-90]{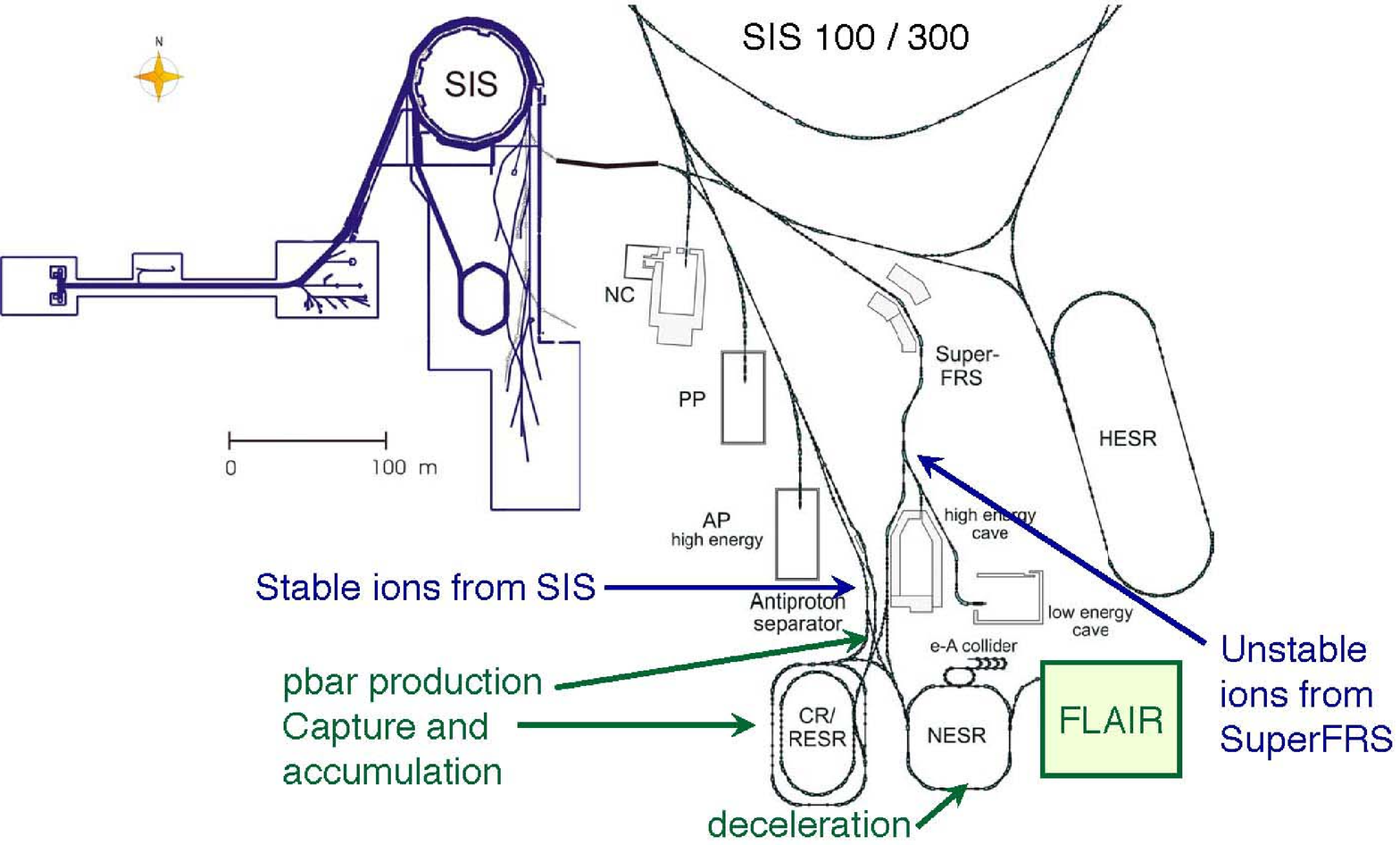}
\caption{FLAIR and the future facility at Darmstadt. SIS: heavy
ion synchrotron, SIS 100: synchrotron with 100 Tm magnetic
rigidity, CR: cooler ring, RESR: recycled experimental storage
ring, NESR: new experimental storage ring. SuperFRS: planned
fragment separator.} \label{fig:FLAIRandFutureFacility}
\end{center}
\end{figure}



The antiproton production method at the future facility \cite{CDR}
(cf. Fig.~\ref{fig:FLAIRandFutureFacility}) is similar to the
scheme used for LEAR at CERN. 10$^8$ antiprotons will be produced
every 5 seconds, will be collected in the CR storage ring,
accumulated in the RESR and transferred to the NESR for further
cooling and deceleration to 30 MeV. From the NESR the antiprotons
will be transferred to the FLAIR facility for further deceleration
in the LSR storage ring, in the electrostatic storage ring USR,
and in the HITRAP facility. Antiproton rates of $\sim 10^6$/s
extracted both fast and slow at energies down to 20 keV  will be
available. The \pbar\  can be either directly stopped in
low-density gas targets using thin windows, or trapped with high
efficiency in charged particle traps. The same rate of antiprotons
can be obtained at rest in HITRAP or extracted at ultra-low
(eV--keV) energies. This scheme gives about a factor 100 more
antiprotons per unit time stopped in gas targets or trapped in ion
traps as compared to the present AD at CERN where no dedicated
accumulation and multi-stage deceleration rings are utilized.
The availability of such beams
will tremendously increase the number of experiments possible at
this facility.



Antiprotons can be extracted either slow or fast from the LSR and
used in several beamlines for e.g. deceleration and cooling in the
HITRAP facility, or for stopping them in low-density gas targets.
In addition, an electrostatic storage ring is foreseen (USR),
which will be used for deceleration, slow and fast extraction, and
for atomic physics experiments with internal targets. The energy
range of 300 - 20 keV (using electron cooling) makes it a unique
tool for many atomic collision experiments which are only possible
in such a low-energy storage ring, where effective intensities
({\em i.~e.} the number of stored particles $N_\mathrm{stored}$
times the revolution frequency $f_\mathrm{rev}$, $R_\mathrm{eff} =
N_\mathrm{stored} f_\mathrm{rev}$) of $R_\mathrm{eff} = 10^{10}$
\pbar /s are reached for in-ring experiments.

\section{Physics program of FLAIR}

\label{sec:physics}


The physics of FLAIR covers a wide range in atomic, nuclear and
particle physics and has potential medical applications. It is
described in detail in the FLAIR letter of intent \cite{FLAIR-LOI}
available from the FLAIR web page. In the following a brief
overview will be given.


\subsection{Precision spectroscopy of antiprotonic atoms and
antihydrogen}

This is the current topic of the Antiproton Decelerator (AD) at
CERN. The main goal here is to study fundamental symmetries and
interactions by providing high-precision data of particle and
antiparticle properties for tests of CPT symmetry and QED
calculations. Antiprotonic atoms have been used for some time to
test CPT symmetry between proton and antiproton properties (for a
review, see \cite{Eades:99}).
The most accurate test of proton/antiproton properties is the
measurement of their cyclotron frequency $\omega_c \propto Q/M$
($Q,M$ denoting charge and mass) by the TRAP collaboration at
LEAR, yielding an accuracy of better than $10^{-10}$
\cite{Gabrielse:99}. Separate CPT limits on $Q$ and $M$ can be set
by combining this measurement with the recent precision laser
spectroscopy of {\em antiprotonic helium} by the ASACUSA
collaboration at the Antiproton Decelerator (AD) of CERN, which
are now at a level of $10^{-8}$ \cite{Hori:03}. These latter
experiment also constitute a sensitive test for three-body QED
calculations. Triggered by the spectroscopy results, the accuracy
of three-body bound-state QED calculations has been tremendously
improved, reaching a level comparable to the experimental
resolution \cite{Korobov:03,Kino:03}.

A further increase by a factor 10 in precision will be attempted
by two-photon laser spectroscopy using a pulse-amplified cw laser.
The pulse-amplification is needed to create enough laser power
density to saturate the transitions over the large stopping
distribution (several cm$^3$) possible at the AD. Increasing the
precision beyond this level will be only possible using pure cw
lasers. Such lasers produce enough power density if they are
focused to typically 1 mm$^2$, like it has been done e.~g. for
positronium spectroscopy \cite{Fee:93a,Fee:93b}. Using a
low-energy cooled antiproton beam as proposed for FLAIR,
antiprotons can be stopped in a region of this size, and a
precision of a few 100 kHz can be expected in the spectroscopy of
antiprotonic helium. This can be equally applied to two-body
systems like protonium or antiprotonic helium ions.

The ultimately highest precision of a CPT test with antiprotonic
atoms is likely to be achieved using {\em antihydrogen}. The
production of large amounts of cold antihydrogen at the AD has
been reported in 2002
\cite{ATHENA-Hbar:02,ATRAP-Hbar:02a,ATRAP-Hbar:02b}, but it is
still expected to take several years until precision spectroscopy
can be performed. After a shutdown in 2005, the AD is expected to
run until 2010, so that initial results on spectroscopy can be
expected at the AD. The ultimate goal is to measure the 1S--2S
two-photon laser transition \cite{ATHENA:96,ATRAP:97} and the
ground state hyperfine splitting \cite{HbarLOI} to accuracies
similar to the ones achieved for hydrogen ($10^{-14}$
\cite{Niering:00} and $10^{-12}$ \cite{Hellwig:70,Essen:71},
resp.), yielding a very sensitive test of the CPT theorem. New
 ideas like the use of a cusp trap for \Hbar\ formation
\cite{MohriEP} are being proposed. To achieve the ultimate
precision, the trapping and laser cooling of neutral antihydrogen
atoms is required. The development of these techniques will surely
take many more years to accomplish.



Once trapped and laser-cooled antihydrogen is available, other
challenging experiments can be performed. Among them is the {\em
gravitation of antimatter} \cite{grav:2003}, which is a long
standing question that has never been answered experimentally,
because in the case of charged particles, gravitational effects
are covered by the many orders of magnitude stronger
electromagnetic interaction. Collisions between antihydrogen and
matter atoms as well as the creation of larger antimatter systems
like \Hbar $^+$ (one antiproton and two positrons, equivalent to
the well known H$^-$ ion) are of big interest for atomic collision
theory.

\subsection{Atomic collision physics}

This field will greatly benefit from the availability of
ultra-slow, cooled antiproton beams in storage rings. This will
enable for the first time ever the detailed study of ionization
processes with antiprotons in kinematically complete experiments.
The energy loss can be investigated at ultra-low energies to
answer open questions about the velocity dependence in this
regime. Antiprotons are best suited for such studies, because
unlike protons their charge is not screened by electrons which
makes the theoretical treatment very difficult. The very short
interaction time of less than femtoseconds for \pbar\ energies
above 1 keV makes antiprotons a perfect and unique tool to study
many-electron dynamics in the strongly correlated, non-linear,
sub-femtosecond time regime, the most interesting and, at the same
time, most challenging domain for theory.

\subsection{Antiprotons as hadronic probes}

In {\em nuclear physics}, the antiproton is used as a hadronic
probe to study the nuclear structure. X-ray spectroscopy of the
low-lying states of \pbar p or other light atoms \cite{Got04}
gives important information on the nucleon-antinucleon interaction
in the low-energy limit, where scattering experiments  cannot
provide precise values. These data are vital for the improvement
of QCD calculations in the low-energy (hence non-perturbative)
region. X-ray spectroscopy of heavy antiprotonic atoms can be used
to obtain information about the density ratio of neutron and
protons at the nuclear periphery, {\em i.~e.} to investigate
neutron halo or skin effects. The PS209 experiment at LEAR has in
this way provided benchmark data for nuclear structure
calculations over a wide range of nuclei \cite{Trzcinska:01}. This
technique is much more sensitive than others like total absorption
cross section measurements, and further systematic measurements
with stable isotope targets will provide a more complete and
systematic picture of the nuclear surface. Since halo effects are
expected to be more pronounced in nuclei with a large neutron
excess which are unstable, the application of this technique to
unstable radioactive ions \cite{Wada-leap03} available at FLAIR
via the SuperFRS will generate important contributions to the
study of the structure of nuclei far from stability.

Concerning the {\em particle physics} point of view, the
elementary {\em antinucleon-nucleon interaction} processes still
present a few unclear aspects \cite{Filippi-leap03}, in spite of
the wealth of physics results produced at LEAR. In particular,
some anomalous effects have been observed close to the
nucleon-antinucleon threshold which are likely related to the
interplay between quark and antiquark degrees of freedom -- that
could be, for instance, responsible of the existence of
quasi-nuclear subthreshold nucleon-antinucleon bound states, long
sought for at LEAR without large success. An effective way to get
a more transparent environment to study the nucleon-antinucleon
interaction dynamics, free from Coulombian contributions and from
a few selected initial states, is the use of {\em antineutrons} as
probes. FLAIR will be able to provide antiproton beams of
intensity and momentum resolution good enough to produce a
continuous antineutron beam, with which new high statistics
measurements of elastic, annihilation and total cross section in
the region  below 100 MeV/$c$, only marginally covered by LEAR,
could be pursued in a relatively short time.

The study of baryon-baryon interactions as a basic tool for
investigations of the strong interaction can be extended to the
hyperon sector, where much less data exist than in the nucleon
sector. Especially few data exist on {\em strangeness S $= -2$}
systems. Stopped antiprotons are very efficient for the production
of S $= -2$ systems via the double strangeness and charge exchange
reaction ($\bar{K^*} , K$) \cite{kil88}. With a sizeable branching
ratio the annihilation of antiprotons results in the production of
a $\bar{K^*}$ ``beam'' which interacts with another nucleon via
$\bar{K^*} N \rightarrow K \Xi$. The momenta of the $\bar{K^*}$
are well matched for the production of slow $\Xi$ particles which
undergo efficient $\Xi N$ interactions. The proposed studies will
result in detailed information of S $= -2$ baryonic and possible
dibaryonic states.

\subsection{Medical applications}

Recently, interest has been shown in the {\em medical application}
of antiprotons for tumor therapy. This comes from the fact the
antiprotons, in addition to depositing energy via their energy
loss like other charged particles, annihilate when stopped in
material. The annihilation produces residual nuclear fragments of
high charge and low energy, which deposit a large biological dose
in the immediate surrounding of the \pbar\ stopping distribution.
Since the cooled low-emittance antiproton beams can be stopped in
a well-defined region, the presumably large energy deposited
locally makes them a suitable tool for tumor therapy. A test
experiment is under way at the AD of CERN \cite{ACEprop} and, if
this effect is confirmed, the method can be extended at FLAIR
where the high-energy antiproton beams (50 -- 300 MeV) needed to
penetrate deep enough into human tissue are available directly
from the NESR.


\section{Conclusions}

FLAIR will be a unique next-generation low-energy antiproton and
ion facility.  Cooled antiprotons down to 20 keV both in storage
rings and extracted will revolutionize low-energy antiproton
physics. Continuously extracted beams at these energies will
enable nuclear and particle physics type experiments currently not
possible at the AD of CERN. The availability of short-lived exotic
nuclei at the future facility at Darmstadt creates synergies by
using antiprotons as hadronic probes for nuclear structure. Using
the same facility, atomic physics experiments with highly charged
ion will be possible both in beams and at rest in HITRAP. If the
letter of intent will be positively evaluated, the FLAIR community
will proceed towards a more detailed design and procurement of the
necessary funding.

\section*{Acknowledgements}

I would like to thank the members of the FLAIR steering committee
and community who contributed to the combined effort to write the
letter of intent and who are working towards the creation of this
new facility.


\end{document}